\documentclass[10pt,journal,twocolumn]{IEEEtran}
\IEEEoverridecommandlockouts
\usepackage{epsfig,cite,bm,algorithm,algorithmic,epstopdf,amsmath,subfigure,multirow,amssymb,amsfonts,amsthm,xcolor,caption,graphicx,textcomp,xcolor, subfigure, makecell, colortbl,utfsym,soul,bbding, floatflt}
\usepackage[inline, shortlabels]{enumitem}
\setlist{itemjoin ={,\enspace},itemjoin* = { and\enspace}}
\def\QEDclosed{\mbox{\rule[0pt]{1.5ex}{1.5ex}}}
\usepackage[T1]{fontenc}
\usepackage{stfloats}
\allowdisplaybreaks[4]

\allowdisplaybreaks[4]
\usepackage{booktabs}

\soulregister{\cite}7
\soulregister{\ref}7

\begin{document}

\title{Hybrid Beamfocusing Design for RSMA-Enabled Near-Field Wideband Communications}

\author{Jiasi Zhou, Chintha Tellambura,~\IEEEmembership{Fellow,~IEEE}

\thanks{Jiasi Zhou is with the School of Medical Information and Engineering, Xuzhou Medical University, Xuzhou, 221004, China, (email: jiasi\_zhou@xzhmu.edu.cn). (\emph{Corresponding author: Jiasi Zhou}).}
\thanks{ Chintha Tellambura is with the Department of Electrical and Computer Engineering, University of Alberta, Edmonton, AB, T6G 2R3, Canada (email: ct4@ualberta.ca).} 
\thanks{This work was supported by the Talented Scientific Research Foundation of Xuzhou Medical University (D2022027).}}
\maketitle
	
\begin{abstract}
Future wireless networks will utilize extremely large-scale antenna arrays (ELAAs) over high-frequency bands, which, however, produce near-field spherical wavefronts and spatial wideband effects. To exploit and mitigate these, this paper proposes a rate-splitting multiple access (RSMA)-enabled transmit scheme for wideband near-field communications (NFC). Our solution leverages true-time-delay (TTD)-based hybrid beamfocusing architectures to mitigate spatial wideband effect and reduce radio frequency chain requirements. The objective is to maximize the minimum rate by jointly optimizing frequency-dependent analog beamfocusing, frequency-independent analog beamfocusing, digital beamfocusing, and common rate allocation. To solve this complicated non-convex problem, we develop a penalty-based iterative algorithm that partitions the variables into three blocks and then employs block coordinate descent (BCD) to optimize each block alternately. This algorithm is further extended to support the sub-connected TTD-based analog beamfocusing architectures. Comprehensive simulation results indicate that our transmit scheme: 1) effectively compensates for spatial wideband effect, addressing a critical challenge in wideband operation; 2) achieves performance comparable to full digital beamfocusing while maintaining lower hardware complexity; 3) achieves substantial performance gains over the other two benchmarks.
\end{abstract} 

\section{Introduction}
With the global commercialization of fifth-generation (5G) networks, sixth-generation (6G) wireless systems have received increasing attention from both academia and industry. According to 3GPP Release 20, 6G is expected to operate over an extended frequency range up to $52.6$~GHz, encompassing millimeter-wave (mmWave) and terahertz (THz) bands. However, signals at these high frequencies suffer from severe path loss, posing a significant challenge for reliable communication. To mitigate this, extremely large-scale antenna arrays (ELAAs) have been proposed. Their deployment, however, extends the Rayleigh distance, causing most devices to operate in the near-field region—where electromagnetic (EM) propagation shifts from far-field plane waves to near-field spherical waves \cite{10559261,10135096}.

Unlike far-field plane wavefronts, near-field spherical wavefronts introduce an extra distance dimension, simultaneously capturing angle and distance information\cite{10220205}. This dual-dimensional spatial resolution empowers unprecedented beamforming precision, concentrating radiated energy onto designated spatial positions rather than merely directed along specific directions\cite{10579914}. This transformative capability, technically termed \emph{beamfocusing}, facilitates finer-grained spatial multiplexing and precise signal enhancement. However, harnessing these benefits calls for thorough overhauls of wireless network architectures to accommodate the new propagation paradigm.

Furthermore, multiple access schemes, governing spectrum sharing and interference management strategies, are critical to meet the ultra-dense connectivity demands. Current near-field communications (NFCs) primarily employ space-division multiple access (SDMA)\cite{zhang2022beam,10436390,10767351} and non-orthogonal multiple access (NOMA)\cite{zuo2023non}. SDMA fully treats interference as noise, while NOMA fully decodes all stronger interference. Both of them lack flexible interference management capability. As a remedy, rate-splitting multiple access (RSMA) distinguishes itself through its superior versatility and robustness in managing interference\cite{mao2018rate}. By adaptively steering the message splitting percentage, RSMA  seamlessly generalizes both SDMA and NOMA as special cases\cite{10038476}. This inherent adaptability enables RSMA to deliver enhanced connectivity, higher spectral efficiency, and better user fairness over SDMA and NOMA\cite{9835151,9991090,10032267}. Due to its potential, RSMA has been considered for NFC in \cite{10798456,singhrsma}, demonstrating its considerable performance gains. Nevertheless, several technical challenges have yet to be addressed in RSMA-enabled NFC, which are summarized below. 

\begin{itemize}
\item \textbf{High Hardware Requirements:} Full digital beamfocusing architectures require a dedicated radio frequency (RF) chain per antenna, incurring prohibitive power consumption and hardware cost\cite{2024Towards}. This issue becomes especially pronounced in NFC, which is usually accompanied by ELAAs\cite{10559261}. In contrast, hybrid analog-digital (HAD) beamfocusing architectures dramatically lower hardware overhead while maintaining a favorable performance-practicality trade-off. As a result, it is particularly well-suited for NFC systems.

\item \textbf{Spatial Wideband Effect:} NFC predominantly operates in high-frequency bands. Such wideband channels exhibit \emph{frequency-dependent} characteristics due to the substantial bandwidth across subcarriers. However, conventional HAD architectures are limited to frequency-independent analog beamfocusing\cite{10659326}. This mismatch causes the beam split effect, where the beam is defocused at the designated location for most subcarriers\cite{10663786}. Consequently, severe performance degradation occurs in wideband NFC. This requires an advanced architecture that supports frequency-aware beamfocusing.
\end{itemize}

Note that RSMA alone cannot eliminate the spatial wideband effect, which degrades the performance of  RSMA-enabled NFC. To our knowledge, the spatial wideband effect in RSMA-enabled NFC has yet to be resolved, which motivates our work. 

\subsection{Related Works}
\subsubsection{Narrowband NFC}
Extensive research has focused on narrowband NFC. For instance, near-field beamfocusing has been studied using full digital, hybrid analog-digital (HAD), and dynamic metasurface antenna architectures, demonstrating significantly higher sum rates than far-field beamforming \cite{zhang2022beam}. It also enhances physical-layer security, particularly when eavesdroppers are aligned with or closer to the BS than legitimate users \cite{10436390}. Recent work \cite{10767351} explores fluid antenna arrays, jointly optimizing beamfocusing vectors and antenna positioning. These studies leverage spatial beamfocusing to suppress inter-user interference \cite{zhang2022beam,10436390,10767351}. However, when many users are scheduled per time slot, interference can degrade performance \cite{9835151}. To mitigate this, a NOMA-based transmit scheme with far-to-near successive interference cancellation (SIC) decoding has been proposed \cite{zuo2023non}, a capability not feasible in far-field NOMA. Together, these advances underscore the increasing significance of NFC.

\subsubsection{Wideband NFC} Recent advances in spatial wideband effect mitigation include algorithmic optimizations and hardware-based compensation methods. The algorithmic approach enhances beamfocusing capability by generating wide beams to achieve flattened beamfocusing gain across the entire bandwidth\cite{10221794,10505154}. However, this approach remains limited by beam split effects due to its reliance on phase shift (PS)-based analog beamfocusing. To overcome this limitation, true-time-delay (TTD)-based hybrid architectures emerge, which can facilitate frequency-dependent analog beams to counteract the spatial wideband effect\cite{10458958,10587118,10659326,10541333}. The authors in \cite{10458958} design frequency-dependent beamfocusing under three different TTD configurations: parallel, serial, and hybrid serial-parallel. A sub-connected TTD architecture is proposed to improve energy efficiency and reduce hardware requirements \cite{10587118,10541333}. Based on this insight, a similar TTD architecture is extended to reconfigurable intelligent surface (RIS)-assisted NFC, where a deep learning-based optimization framework is proposed to maximize spectral efficiency\cite{10659326}.

\subsubsection{RSMA-enabled narrowband NFC} RSMA exhibits superior anti-interference benefits, but remains largely underexplored in NFC, except for\cite{10414053,10798456,singhrsma,10906379,zhou2025crb,zhou2024hybrid,11071287}. Specifically, the preconfigured spatial beamfocusing is repurposed to support additional communication users \cite{10798456} or sense an extra near-field target \cite{10906379}. The authors in \cite{11071287} utilize RSMA to counter interference in NFC with imperfect channel state information (CSI) and SIC. The transmit scheme is extended for RIS-enabled NFC\cite{singhrsma}. However, these implementations depend exclusively on full digital beamfocusing architectures, which incur substantial hardware complexity for practical deployments. To tackle this fundamental limitation, the HAD-based beamfocusing architecture is developed for mixed near-field and far-field communications\cite{10414053}. Furthermore, similar beamfocusing architectures have been extended to RSMA-enabled near-field integrated sensing and communication (ISAC) networks \cite{zhou2024hybrid,zhou2025crb}. However, these efforts can only generate frequency-independent analog beamfocusing, limiting their effectiveness in wideband operation. 

\subsection{Contributions}
To address this critical gap, this paper proposes TTD-based hybrid beamfocusing architectures for  RSMA-enabled wideband NFC and investigates the corresponding hybrid beamfocusing design. Table \ref{Table I}  compares the existing NFC studies, highlighting the novelty of our work. Our contributions are outlined below:
\begin{table*}[h]
	\caption{Our contributions in contrast to the existing NFC}
	\begin{center}\label{Table I}
		\begin{tabular}{|c||c|c|c|c|c|c|c|} 
			\hline
&\cite{zhang2022beam,10436390,zuo2023non} &\cite{10767351}&\cite{10221794,10505154}&\cite{10458958,10587118,10659326,10541333}&\cite{10798456,10906379,singhrsma}&\cite{10414053,zhou2024hybrid, zhou2025crb}&\makecell*[c]{\bf{Our work}}\\
                \hline 
           	\makecell*[c]{\bf{RSMA}} & \XSolidBrush & \XSolidBrush&\XSolidBrush &\XSolidBrush &\Checkmark&\Checkmark&\Checkmark\\
                \hline 
           	\makecell*[c]{\bf{Spatial wideband effect }} & \XSolidBrush& \XSolidBrush&\Checkmark &\Checkmark &\XSolidBrush&\XSolidBrush&\Checkmark\\
	          \hline 
           	\makecell*[c]{\bf{Hybrid beamfocusing}} & \Checkmark&\XSolidBrush & \XSolidBrush&\Checkmark &\XSolidBrush &\Checkmark &\Checkmark\\
	          \hline
		\end{tabular}
	\end{center}
\end{table*}
\begin{itemize}
\item \textbf{Transmit Scheme:}
We propose a fully-connected TTD-based hybrid beamfocusing architecture tailored for RSMA-enabled wideband NFC systems. The RSMA framework offers robust and flexible interference management, while the TTD-based hybrid beamfocusing effectively mitigates the spatial wideband effect and alleviates RF hardware constraints. Our primary goal is to maximize the minimum achievable rate across all users. To this end, we jointly optimize key system components: frequency-dependent and frequency-independent analog beamfocusing, digital beamfocusing, and common rate allocation.

\item \textbf{Algorithm Framework:} We propose a penalty-based iterative algorithm for the fully-connected TTD-based hybrid beamfocusing design, which introduces auxiliary matrices to decouple the formulated optimization problem. The algorithm  employs the block coordinate descent (BCD) methodology that alternately optimizes three distinct components:
\begin{enumerate}
\item \emph{Auxiliary variables and common rate allocation optimization}: We propose an iterative optimization algorithm based on the minorization-maximization (MM) framework. A key challenge in the MM framework lies in constructing effective surrogate functions for complex objectives. Our solution builds concave quadratic surrogates for logarithmic transmit rates. These surrogates satisfy the minorization property and gradient consistency, exhibiting superior convergence performance\cite{mairal2013optimization}.
\item \emph{Analog beamfocusing optimization:} We develop an alternating optimization algorithm. It iteratively updates the frequency-dependent analog beamfocusing through a one-dimensional search and computes the frequency-independent counterpart via a closed-form solution.
\item \emph{Digital beamfocusing optimization:} The optimal digital beamfocusing is obtained via a closed-form expression.  
\end{enumerate}

\item \textbf{Algorithm Extension:} To reduce the number of TTDs and their maximum delay requirement, we propose a sub-connected TTD-based hybrid beamfocusing architecture. Furthermore, we extend the proposed penalty-based iterative algorithm to support such configurations.

\item \textbf{Performance Insights:} Extensive simulations validate the efficiency of our proposed transmit scheme, underscoring three undeniable advantages over four baselines: 1) effective mitigation of the spatial wideband effect, addressing a critical challenge in wideband operation; 2) near full digital beamfocusing performance with reduced hardware complexity; 3) substantial performance gains over SDMA and far-field communications.
\end{itemize}

\emph{Organization:}  The remainder of this paper is organized as follows.  Section \ref{Section II} presents the signal model and formulates a max-min rate optimization problem. Section \ref{Section III} elaborates on the penalty-based iterative algorithm and analyzes its properties. Section \ref{Section IV} extends the proposed algorithm to handle the sub-connected TDD-based hybrid beamfocusing architecture. Section \ref{Section V} provides the simulation results. Finally, this paper is concluded in Section \ref{Section VI}.

\emph{Notations:} Boldface upper-case letters, boldface lower-case letters, and calligraphy letters denote matrices, vectors, and sets, respectively. The complex matrix space of $N\times K$ dimensions is denoted by $\mathbb{C}^{N\times K}$. Superscripts ${(\bullet)}^T$ and ${(\bullet)}^H$ represent the transpose and Hermitian transpose, respectively. $\text{Re}\left( \bullet\right)$ and $\mathbb{E}\left[\bullet\right]$ 
 denote the real part and statistical expectation, respectively. The Frobenius norm of matrix $\mathbf{X}$ is denoted by $||\mathbf{X}||_F$. For matrix $\mathbf{X}$, $\mathbf{X}\left(i:j,:\right)$  represent a sub-matrix composed of the rows from the $i$-th to $j$-th. $\text{diag}\left( \bullet\right)$  and $\text{blkdiag}\left( \bullet\right)$ denote the diagonal and block diagonal operations, respectively. The operator $\angle$ denotes the phase of a complex value. $\mathcal{CN}(\mu, \sigma^2)$ denotes a complex Gaussian of mean $\mu$ and variance $\sigma^2$.

\section{System model and problem formulation}\label{Section II}
We consider an RSMA-enabled wideband NFC system, where a BS equipped with an $N$-antenna uniform linear array (ULA) serves $K$ single-antenna near-field users. The ULA has an inter-element spacing of $d$, yielding a Rayleigh distance of $d_r=\frac{2D^2}{\lambda_c}$, where $D=(N-1)d$ and $\lambda_c$ are antenna aperture and signal wavelength of the central carrier frequency $f_c$, respectively. The set of users is indexed by $\mathcal{K}=\{1,\dots,K\}$. To combat inter-symbol interference in wideband transmission, we adopt orthogonal frequency division multiplexing (OFDM). The total system bandwidth $B$ is uniformly partitioned into $M$ subcarriers, with the $m$-th subcarrier's central frequency given by $f_m=f_c+\frac{B(2m-1-M)}{2M}$ for $\forall m\in\mathcal{M}=\{1,\dots,M\}$.

\subsection{Near-field wideband channel models}
\begin{figure}[tbp]
\centering
\includegraphics[scale=0.6]{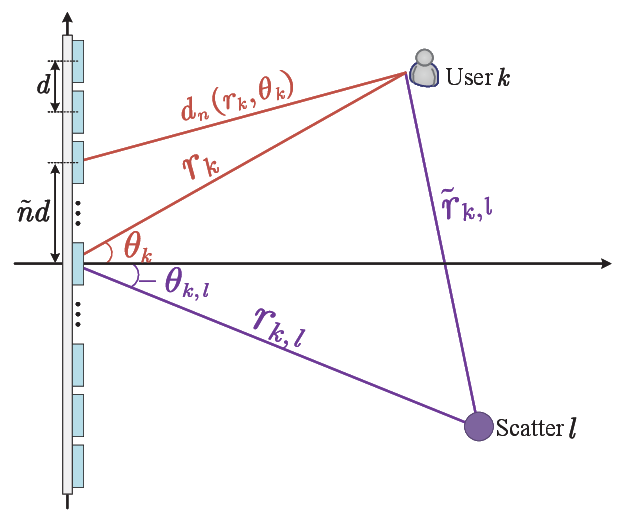}
\caption{Illustration of the multi-path near-field channel model.}
\label{fig:coordinate}
\end{figure}  
As illustrated in Fig.~\ref{fig:coordinate}, we consider a multi-path near-field wideband channel model, comprising a line-of-sight (LoS) path and $L_k$ non-LoS (NLoS) paths incurred by $L_k$ scatters.  For our coordinate system, we position the ULA midpoint at the origin with the array aligned along the $y$-axis. Under this configuration, the coordinate of the $n$-th antenna is $\mathbf{s}_n=\left(0, \tilde nd\right)$, where $n\in\mathcal{N}=\{1,\dots,N\}$ and $\tilde n=\frac{2n-N-1}{2}$. Let the $k$-th user located at $r_k$ distance and $\theta_k$ angle from the coordinate origin, so its coordinate is $\mathbf{r}_k=\left(r_k\cos\theta_k,r_k\sin\theta_k\right)$. The distance between the $n$-th antenna and the $k$-th user can be given by
\begin{align}
d_{n}\left(r_k,\theta_k\right)=&||\mathbf{r}_k-\mathbf{s}_n|| = \sqrt{r^2_k+(\tilde nd)^2-2\tilde ndr_k\sin\theta_k}\notag\\
\approx& r_k-\tilde nd\sin\theta_k+(\tilde nd)^2\cos^2\theta_k/2r_k.
\end{align}
The approximation employs a second-order Taylor expansion to capture spherical wavefront characteristics accurately\cite{10579914}. Furthermore, within the Fresnel region, channel gains remain approximately uniform across all antenna elements\cite{10135096}. Consequently, the channel between the $n$-th antenna and the $k$-th user at the $m$-th subcarrier can be modelled as
\begin{align}
\tilde h_{n,k,m}\left(r_k,\theta_k,f_m\right)= &\tilde\beta_{k,m} e^{-j\frac{2\pi f_m}{c}d_{n}\left(r_k,\theta_k\right)}\notag\\\approx&\beta_{k,m} e^{j\frac{2\pi f_m}{c}\delta_{n}\left(r_k,\theta_k\right)},
\label{Single_Channel}
\end{align}
where $\delta_{n}\left(r_k,\theta_k\right)=\tilde nd\sin\theta_k-(\tilde nd)^2\cos^2\theta_k/2r_k$. Here, $\tilde\beta_{k,m}=\frac{c}{4\pi f_mr_k}$ and $\beta_{k,m}=\tilde\beta_{k,m} e^{-j\frac{2\pi f_m}{c}r_k}$ respectively denote the pathloss and complex channel gain with $c$ being the speed of light. Aggregating all antenna elements, the LoS channel vector is
\begin{align}
\tilde{\mathbf{h}}_{k,m}&\approx \beta_{k,m}\left[e^{j\frac{2\pi f_m}{c}\delta_{1}\left(r_k,\theta_k\right)},\dots,e^{j\frac{2\pi f_m}{c}\delta_{N}\left(r_k,\theta_k\right)}\right]^T\notag\\&=\beta_{k,m}\mathbf{a}\left(r_k,\theta_k,f_m\right).
\label{Channel}
\end{align}
where $\left(r_k,\theta_k,f_m\right)$ in $\tilde{\mathbf{h}}_{k,m}$ is omitted for brevity and $\mathbf{a}\left(r_k,\theta_k,f_m\right)$ is the near-field array response vector.
Let $r_{k,l}$ and $\theta_{k,l}$ represent the distance and angle of the $l$-th scatter associated with user $k$. Incorporating $L_k$ scatters, the overall channel $\mathbf{h}_{k,m}\in\mathbb{C}^{N\times 1}$ can be characterized as
\begin{equation}
\mathbf{h}_{k,m}= \beta_{k,m}\mathbf{a}\left(r_k,\theta_k,f_m\right) + \sum_{l=1}^{L_k}\beta_{k,m,l}\mathbf{a}\left(r_{k,l},\theta_{k,l},f_m\right),
\label{Overall_Channel}
\end{equation}
where $\beta_{k,m,l}=\tilde\beta_{k,m,l} e^{-j\frac{2\pi f_m}{c}\left(r_{k,l}+\tilde r_{k,l}\right)}$ accounts for the complex channel gains of the $l$-th NLoS components, and $\tilde r_{k,l}$ is the distance between the $k$-th user and the $l$-th scatter associated with user $k$. 

\subsection{TTD-based hybrid beamfocusing for RSMA systems }
\begin{figure*}[tbp]
\centering
\includegraphics[scale=0.6]{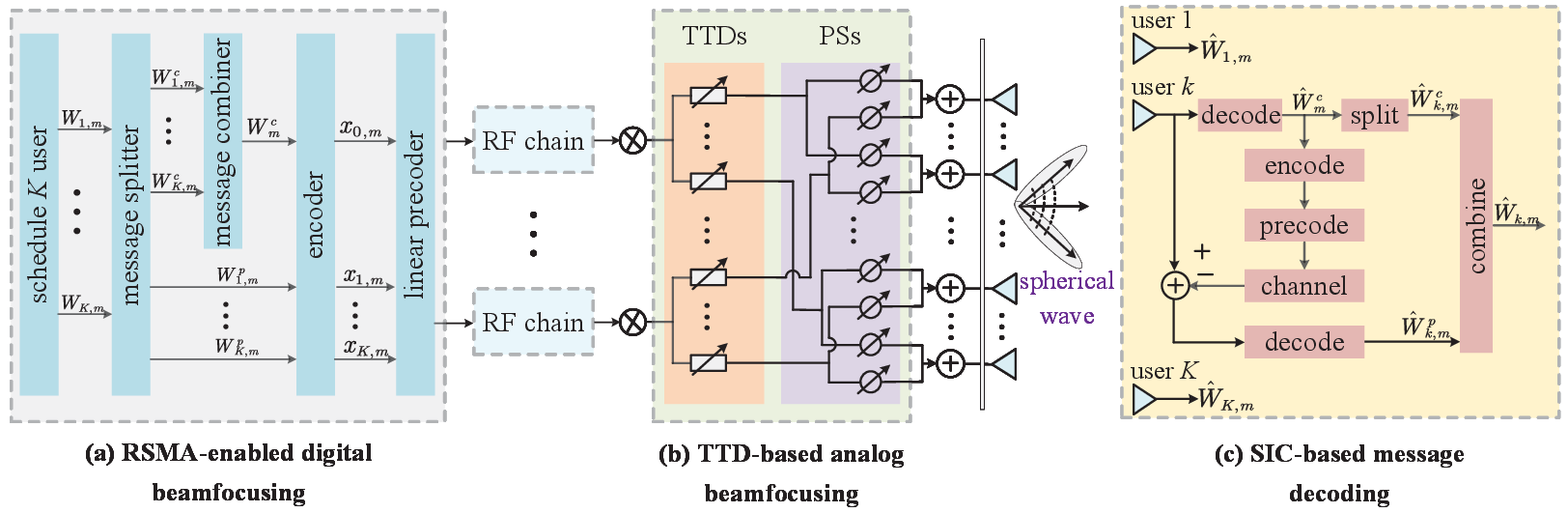}
\caption{RSMA-enabled Wideband NFC systems with TTD-based hybrid beamfocusing architecture.}
\label{fig:system}
\end{figure*}  
\subsubsection{RSMA-enabled digital beamfocusing} As per Fig.~\ref{fig:system}(a), in the RSMA system,   the message $W_{k,m}$  for the $k$-th user at subcarrier $m$ is split into a common part $W^c_{k,m}$ and a private part $W^p_{k,m}$ for $\forall k\in\mathcal{K}$ and $\forall m\in\mathcal{M}$. All common parts $\left\{W^c_{1,m},\dots, W^c_{K,m}\right\}$ at subcarrier $m$ are combined and encoded into one common stream $x_{0,m}$ using a public coodbook. The private parts $\left\{W^p_{1,m},\dots,W^p_{K,m}\right\}$ are independently encoded into user-specific  streams $\left\{x_{1,m},\dots,x_{K,m}\right\}$. All streams are mutually independent and meet the unit power constraint $\mathbb{E}|x_{k,m}|^2=1$ for $\forall k \in\tilde{\mathcal{K}}=\{0,1,\dots, K\}$ and $\forall m\in\mathcal{M}$. These streams at subcarrier $m$ are then precoded by the digital beamfocusing matrix $\mathbf{W}_m=\left[\mathbf{w}_{0,m},\mathbf{w}_{1,m},\dots,\mathbf{w}_{K,m}\right]\in\mathbb{C}^{A\times (K+1)}$, where $A$ denotes the number of RF chains and $K+1\leq A\leq N$. In $\mathbf{W}_m$, $\mathbf{w}_{0,m}\in\mathbb{C}^{A\times 1}$ and $\mathbf{w}_{k,m}\in\mathbb{C}^{A\times 1}$ are designed for the common stream and the $k$-th private stream at subcarrier $m$, respecitvely. The resulting digital signal at subcarrier $m$ is $\tilde{\mathbf{x}}_m=\mathbf{w}_{0,m}x_{0,m}+\sum^{K}_{k=1}\mathbf{w}_{k,m}x_{k,m}$, which is subsequently processed by the TDD-based analog beamfocusing for transmission. These basic assumptions are standard in the literature\cite{mao2018rate}.  

\subsubsection{TTD-based analog beamfocusing}
To mitigate the spatial wideband effect, we propose a novel analog beamfocusing architecture incorporating TTD units. As per Fig.~\ref{fig:system}(b), an additional fully-connected TTD network is inserted between the PS array and RF chains. In this configuration, each RF chain connects to $Q$ TTD elements, enabling frequency-dependent analog beamfocusing. Consequently, the frequency-dependent analog beamfocusing matrix $\mathbf{T}_m\in\mathbb{C}^{AQ\times A}$ for subcarrier $m$ can be expressed as  
\begin{equation}\label{dependent}
\mathbf{T}_m = \text{blkdiag}\left(e^{-j2\pi f_m\mathbf{t}_1},\dots,e^{-j2\pi f_m\mathbf{t}_{A}}\right),
\end{equation}
where $\mathbf{t}_a=\left[t_{a,1},\dots,t_{a,Q}\right]^T\in\mathbb{C}^{Q\times 1}$ denotes the time-delay vector implemented by the TDDs connected to the $a$-th RF chain, with $\forall a \in\mathcal{A}=\{1,\dots, A\}$. To accommodate practical hardware constraints on complexity and physical size, each TTD element must satisfy a maximum delay limitation, i.e., $t_{a,q}\in\left[0,t_{\text{max}}\right]$
for $\forall a\in\mathcal{A}$ and $\forall q\in\mathcal{Q}=\{1,\dots,Q\}$. In this architecture, each TTD drives $N/Q$ antennas through a dedicated PS network. Thus, the frequency-independent analog beamfocusing matrix $\mathbf{F}\in\mathbb{C}^{N\times AQ}$ is given by
\begin{subequations}\label{independent}
\begin{align}
& \mathbf{F} =\left[\mathbf{F}_1,\dots,\mathbf{F}_A\right],\\
&\mathbf{F}_a= \text{blkdiag}\left(\mathbf{f}_{a,1},\dots,\mathbf{f}_{a,Q}\right),~\forall a\in\mathcal{A},
\end{align}
\end{subequations}
where $\mathbf{F}_a\in\mathbb{C}^{N\times Q}$ and $\mathbf{f}_{a,q}\in\mathbb{C}^{N/Q\times 1}$ represent the PS-based analog beamfocusing connected the $a$-th RF chain and the $q$-th TTD of this chain, respectively. Due to the inherent hardware limitations of PS implementations, all non-zero entries in $\mathbf{F}$ must meet the unit-modulus constraint, i.e., $\left|\mathbf{f}_{a,q,i}\right|=1$ for $\forall a\in\mathcal{A}$ and $\forall q\in\mathcal{Q}$, where $\mathbf{f}_{a,q,i}$ is the $i$-th element of vector $\mathbf{f}_{a,q}$ for $1\leq i\leq N/Q$.

\subsubsection{RSMA-enabled message decoding and transmit rate} After precoding with fully-connected TTD-based analog beamfocusing $\mathbf{FT}_m$, the transmitted signal at subcarrier $m$ can be expressed as $\mathbf{x}_m=\mathbf{FT}_m\tilde{\mathbf{x}}_m$.
Then, at the $k$-th user end, the received signal at the $m$-th subcarrier can be given by
\begin{equation}
y_{k,m}=\mathbf{h}^H_{k,m}\mathbf{FT}_m\left(\mathbf{w}_{0,m}x_{0,m}+\sum^{K}_{k=1}\mathbf{w}_{k,m}x_{k,m}\right)+n_{k,m},
\end{equation}
where $n_{k,m}\sim \mathcal{CN}\left(0,\sigma^2_{k,m}\right)$ denotes additional white Gaussian noise (AWGN) term. The received power can be derived by equation (\ref{equ:rece_power}) as shown at the top of the next page~\cite{7555358}.
\begin{table*}[th]
\hrule
\begin{equation}
T^c_{k,m}=\overbrace{{\left|\mathbf{h}^H_{k,m}\mathbf{FT}_m\mathbf{w}_{0,m}\right|}^2}^{S^c_{k,m}}+\underbrace{\overbrace{{\left|\mathbf{h}^H_{k,m}\mathbf{FT}_m\mathbf{w}_{k,m}\right|}^2}^{S^p_{{k,m}}}+\overbrace{\sum_{j=1,j\neq k}^{K}\left|\mathbf{h}^H_{k,m}\mathbf{FT}_m\mathbf{w}_{j,m}\right|^2+\sigma^2_{k,m}}^{I^p_{{k,m}}}}_{I^c_{k,m}=T^p_{k,m}}.
\label{equ:rece_power}
\end{equation}
\hrule
\end{table*}

As depicted in Fig.~\ref{fig:system}(c), message decoding process consists of two key stages. First, each receiver performs SIC to decode and remove the common stream by treating private streams as noise. Subsequently, each receiver detects its desired private stream by treating the remaining private streams as noise. Therefore, the achievable rates of decoding common and private streams at the $m$-th subcarrier are respectively given by 
\begin{equation}
R^c_{k,m}=\log\left(1 + \gamma^c_{{k,m}}\right)~\text{and}~R^p_{k,m}=\log\left(1 + \gamma^p_{{k,m}}\right),
\end{equation}
where $\gamma^c_{k,m}=S^c_{k,m}({I^{c}_{{k,m}}})^{-1}$  and $\gamma^p_{k,m}=S^p_{k,m}({I^{p}_{{k,m}}})^{-1}$ are the corresponding signal-to-interference-plus-noise ratios (SINRs). Notably, since the common stream is intended by all users, it is necessary to ensure that all users can successfully decode the common stream. As a result, the common rate at the $m$-th subcarrier shall not exceed $R^c_m= \min_{\forall k}R^c_{k,m}$. Furthermore, since all users share the common rate,  we have $R^c_m=\sum_{k=1}^{K}C^c_{k,m}$, where $C^c_{k,m}$ is the portion of the common rate transmitting $W^c_{k,m}$. As such, the total transmit rate for user $k$ over all subcarriers is $R_k=\sum^{M}_{m=1} R_{k,m}$, where $R_{k,m}=C^c_{k,m}+R^p_{k,m}$ is the achievable rate at the $m$-th subcarrier.

\subsection{Problem formulation}
This paper aims to maximize the minimum rate among all users by jointly optimizing frequency-dependent analog beamfocusing, frequency-independent analog beamfocusing, digital beamfocusing, and common rate allocation. This problem is formulated as
\begin{subequations}\label{linear_p}
	\begin{align}
&\max_{\mathbf{F},\{\mathbf{T}_m\},\{\mathbf{W}_m\},\{C^c_{k,m}\} } \min_{\forall k} R_k,\label{ob_a}\\
	\text{s.t.}~
	&||\mathbf{FT}_m\mathbf{W}_m||^2_F\leq P_{\text{th}},~\forall m,\label{ob_b}\\
 &\sum_{k=1}^{K}C^c_{k,m} \leq R^c_m,~\forall m,\label{ob_c}\\
 &C^c_{k,m} \geq 0, \quad \forall k,\forall m,\label{ob_d}\\
  &|\mathbf{f}_{a,q,i}|=1,~\forall a, \forall q, \forall i,\label{ob_e}\\  
  &t_{a,q}\in\left[0,t_{\text{max}}\right],~\forall a, \forall q,\label{ob_f}
	\end{align}
\end{subequations}
where $P_{\text{th}}$ is the maximum transmit power threshold for each subcarrier. (\ref{ob_b}) limits the transmit power per subcarrier.  (\ref{ob_c}) and (\ref{ob_d}) are the common rate allocation constraints. (\ref{ob_e}) accounts for the unit-modulus constraint of the frequency-independent analog beamfocusing while (\ref{ob_f}) is the hardware constraint of the frequency-dependent analog beamfocusing.

Solving problem (\ref{linear_p}) exhibits three technical challenges. First, the inherent non-convexity and non-smoothness in the objective function and transmit rate make it difficult to solve in the primal and dual domains due to the unknown duality gap. Second, the digital beamfocusing, frequency-independent beamfocusing, and frequency-dependent analog beamfoucusing are highly coupled, complicating the hybrid beamfocusing design. Third, the unit-modulus constraint (\ref{ob_e}) compounds the optimization difficulties. As a result, there is no generic solution to the problem (\ref{linear_p}). Additionally, the global optimal solution appears elusive.

\section{Algorithm design and properties analysis}\label{Section III}
This section develops a penalty-based iterative algorithm, which introduces auxiliary matrices to decouple the formulated non-convex problem. The algorithm partitions the variables into three distinct blocks and leverages block coordinate descent (BCD) to alternately optimize each block. Furthermore, several critical properties of the proposed algorithm, including convergence and complexity, are discussed.

To proceed, we introduce a series of auxiliary matrices $\mathbf{P}_m=\left[\mathbf{p}_{0,m},\dots,\mathbf{p}_{K,m}\right]\in\mathbb{C}^{N\times (K+1)}$ for $\forall m$, which meets $\mathbf{p}_{k,m}=\mathbf{FT}_m\mathbf{w}_{k,m}$ for $\forall k\in\mathcal{\tilde{K}}$. 
\begin{table*}[th]
\hrule
\begin{equation}
T^c_{k,m}=\overbrace{{\left|\mathbf{h}^H_{k,m}\mathbf{p}_{0,m}\right|}^2}^{S^c_{k,m}}+\underbrace{\overbrace{{\left|\mathbf{h}^H_{k,m}\mathbf{p}_{k,m}\right|}^2}^{S^p_{{k,m}}}+\overbrace{\sum_{j=1,j\neq k}^{K}\left|\mathbf{h}^H_{k,m}\mathbf{p}_{j,m}\right|^2+\sigma^2_{k,m}}^{I^p_{{k,m}}}}_{I^c_{k,m}=T^p_{k,m}},
\label{equ:rece_power3}
\end{equation}
\hrule
\end{table*}
Moreover, to attack the non-smoothness due to the minimum operator in the objective function (10a), we introduce a non-negative auxiliary variable $R_\text{r}$. Problem (\ref{linear_p}) can then be reformulated as 
\begin{subequations}\label{linear_p2}
	\begin{align}
&\max_{\mathcal{V}} R_\text{r},\label{ob_a2}\\
	\text{s.t.}~
	&\mathbf{P}_m=\mathbf{FT}_m\mathbf{W}_m,~\forall m,\label{ob_b2}\\
 &||\mathbf{P}_m||^2_F\leq P_{\text{th}},~\forall m,\label{ob_c2}\\
 &R_k\geq R_\text{r},~\forall k,\label{ob_d2}\\
 &\mbox{ (\ref{ob_c}) -- (\ref{ob_f})}.\label{ob_e2}
	\end{align}
\end{subequations}
where set $\mathcal{V}=\left\{\{\mathbf{P}_m\},\mathbf{F},\{\mathbf{T}_m\},\{\mathbf{W}_m\},\{C^c_{k,m}\}, R_\text{r} \right\}$ refers to all optimization variables. Note that the received power by the $k$-th user in computing $R^c_{k,m}$ and $R^p_{k,m}$ needs to be modified to equation (\ref{equ:rece_power3}) as shown at the top of the next page~\cite{7555358}. However, the introduced equality constraint (\ref{ob_b2}) hinders the hybrid beamfocusing design. To tackle this challenge, we adopt the penalty method, where the equality constraint is moved into the objective function\footnote{Another similar alternative is the penalty dual decomposition (PDD) method\cite{9120361}, which introduces the Lagrangian dual variable and penalty factor to create a double-loop iterative problem. The inner loop solves the augmented Lagrangian problem while the outer loop updates the Lagrangian dual matrix and penalty factor. These two methods share the same underlying principle and follow nearly identical solution processes. }. Consequently, the resultant new problem is
\begin{subequations}\label{linear_p3}
	\begin{align}
&\max_{\mathcal{V}} R_\text{r} -\frac{1}{\rho}\sum^{M}_{m=1}||\mathbf{P}_m-\mathbf{FT}_m\mathbf{W}_m||^2_F,\label{ob_a3}\\
	\text{s.t.}~
 &\mbox{ (\ref{ob_c}) -- (\ref{ob_f}), (\ref{ob_c2}), (\ref{ob_d2})}.\label{ob_b3}
	\end{align}
\end{subequations}
where $\rho>0$ is the penalty parameter. In particular, $\rho\to 0$ ensures that the penalty term equals exactly zero, yielding a feasible solution to problem (\ref{linear_p2}). However, to achieve optimal performance, the penalty factor is initially set to a sufficiently large value, enabling effective maximization of the original objective function. Subsequently,  it is gradually reduced to drive the penalty term toward zero, thereby enforcing the equality constraint. This approach optimizes the hybrid beamfocusing sufficiently close to the optimal full digital beamfocusing through the penalty method. The resulting solution provides a theoretical performance upper bound for hybrid beamfocusing architecture. Given the known penalty factor, the problem (\ref{linear_p3}) remains challenging to solve optimally. To tackle this issue, we partition the optimization variables $\mathcal{V}$ into three blocks, namely, $\left\{\{\mathbf{P}_m\},\{C^c_{k,m}\}, R_\text{r} \right\}$ , $\left\{\mathbf{F},\{\mathbf{T}_m\}\right\}$, and $\left\{\mathbf{W}_m\right\}$. Then, the BCD approach is employed to alternately optimize each block, with the detailed procedure provided next.

\subsection{Subproblem w.r.t. $\{\mathbf{P}_m\}$, $\{C^c_{k,m}\}$, and $R_\text{r}$ } 
Given the fixed $\mathbf{F}$, $\{\mathbf{T}_m\}$, and $\left\{\mathbf{W}_m\right\}$, the optimization problem is given by
\begin{subequations}\label{linear_p4}
	\begin{align}
&\max_{\{\mathbf{P}_m\}, \{C^c_{k,m}\}} R_\text{r} -\frac{1}{\rho}\sum^{M}_{m=1}||\mathbf{P}_m-\mathbf{FT}_m\mathbf{W}_m||^2_F,\label{ob_a4}\\
	\text{s.t.}~
  &\mbox{ (\ref{ob_c}), (\ref{ob_d}), (\ref{ob_c2}), (\ref{ob_d2})}.\label{ob_b4}
	\end{align}
\end{subequations}
Problem (\ref{linear_p4}) involves fractional SINR and coupled fully-digital beamfocusing, complicating the solution process. To tackle this problem, we leverage the MM algorithmic framework, which iteratively optimizes a tractable surrogate function that approximates the original complex objectives. However, the effectiveness of this framework critically depends on constructing accurate and easily optimized surrogates. Herein, we build lower-bounded concave quadratic surrogates to approximate logarithmic transmit rates. Specifically, the concave quadratic surrogate for $R^c_{k,m}$ and $R^p_{k,m}$ are respectively given by
\begin{subequations}\label{Overall_SCA}
\begin{align}
&f^c_{k,m}\left(\mathbf{P}_m\right)= \sum_{j=0}^{K}\mathbf{p}^H_{j,m}\mathbf{X}^c_{k,m}\mathbf{p}_{j,m}+2\text{Re}\left(\mathbf{y}^c_{k,m}\mathbf{p}_{0,m}\right)+ z^c_{k,m},
\label{SCA_1}\\
&f^p_{k,m}\left(\mathbf{P}_m\right)= \sum_{j=1}^{K}\mathbf{p}^H_{j,m}\mathbf{X}^p_{k,m}\mathbf{p}_{j,m}+2\text{Re}\left(\mathbf{y}^p_{k,m}\mathbf{p}_{k,m}\right) + z^p_{k,m},
\label{SCA_2}
\end{align}
\end{subequations}
where 
\begin{align}
\mathbf{X}^\tau_{k,m}&=-\frac{1}{\ln 2}\mathbf{h}_{k,m}\tilde{\mathbf{u}}^\tau_{k,m} \left(\tilde v^{\tau}_{k,m}\right)^{-1}\left(\tilde{\mathbf{u}}^\tau_{k,m}\right)^H\mathbf{h}^H_{k,m},\notag\\
\mathbf{y}^\tau_{k,m}&=\frac{1}{\ln 2}\left(\tilde v^{\tau}_{k,m}\right)^{-1}\left(\tilde{\mathbf{u}}^\tau_{k,m}\right)^H\mathbf{h}^H_{k,m},\\
z^\tau_{k,m}&=-\frac{1}{\ln 2}\left(\tilde v^{\tau}_{k,m}\right)^{-1}\left(\sigma^2_{k,m}\left(\tilde{\mathbf{u}}^\tau_{k,m}\right)^H\tilde{\mathbf{u}}^\tau_{k,m}+1\right)\notag\\&\quad+\frac{1}{\ln 2}-\log \tilde{v}^\tau_{k,m},\notag
\end{align}
with $\forall\tau\in\{c,p\}$ and 
\begin{align}\label{Auxiliary_1}
&\tilde{\mathbf{u}}^c_{k,m}=\left(\sum_{j=0}^{K}\mathbf{h}^H_{k,m}\tilde{\mathbf{p}}_{j,m}\tilde{\mathbf{p}}^H_{j,m}\mathbf{h}_{k,m}+\sigma^2_{k,m}\right)^{-1}\mathbf{h}^H_{k,m}\tilde{\mathbf{p}}_{0,m},\notag\\
&\tilde v^c_{k,m}=1-\left(\tilde{\mathbf{u}}^c_{k,m}\right)^H\mathbf{h}^H_{k,m}\tilde{\mathbf{p}}_{0,m},\\
&\tilde{\mathbf{u}}^p_{k,m}=\left(\sum_{j=1}^{K}\mathbf{h}^H_{k,m}\tilde{\mathbf{p}}_{j,m}\tilde{\mathbf{p}}^H_{j,m}\mathbf{h}_{k,m}+\sigma^2_{k,m}\right)^{-1}\mathbf{h}^H_{k,m}\tilde{\mathbf{p}}_{k,m},\notag\\
&\tilde v^p_{k,m}=1-\left(\tilde{\mathbf{u}}^p_{k,m}\right)^H\mathbf{h}^H_{k,m}\tilde{\mathbf{p}}_{k,m}.\notag
\end{align}
In equation (\ref{Auxiliary_1}), $\tilde{\mathbf{p}}_{k,m}$ is the last iteration's updated fully-digital beamfocusing for the $k$-th stream at subcarrier $m$ for $\forall k\in\tilde{\mathcal{K}}$. The theoretical analysis in \cite{mairal2013optimization} demonstrates that surrogate functions satisfying minorization and gradient consistency conditions exhibit good convergence performance. Claim 1 shows that the constructed surrogate $f^\tau_{k,m}\left(\mathbf{P}_m\right)$ meets these two fundamental characteristics.

{\bf{Claim 1:}} The defined surrogate function $f^\tau_{k,m}\left(\mathbf{P}_m\right)$ possesses minorization property and gradient consistency:
\begin{enumerate}
    \item \textbf{Minorization property:} $f^\tau_{k,m}\left(\mathbf{P}_m\right)$ strictly lower-bounds the original logarithmic rate function, i.e., $\log\left(1 + \gamma^\tau_{{k,m}}\right)\geq f^\tau_{k,m}\left(\mathbf{P}_m\right)$ with equality occurs when $\mathbf{p}_{k,m}=\tilde{\mathbf{p}}_{k,m}$ for $\forall k$ and $\forall m$. 
    \item \textbf{Gradient consistency:} $f^\tau_{k,m}\left(\mathbf{P}_m\right)$ maintains first-order consistency with the original function at the expansion point $\tilde{\mathbf{p}}_{k,m}$, i.e.,
    \begin{equation}
    \frac{\partial{f^\tau_{k,m}\left(\mathbf{P}_m\right)}}{\partial{\mathbf{p}_{k,m}}}\Bigg|_{\mathbf{p}_{k,m} = \tilde{\mathbf{p}}_{k,m}}=\frac{\partial{\log\left(1 + \gamma^\tau_{{k,m}}\right)}}{\partial{\mathbf{p}_{k,m}}}\Bigg|_{\mathbf{p}_{k,m} = \tilde{\mathbf{p}}_{k,m}}.
    \end{equation}
\end{enumerate}
\emph{Proof:} Please see Appendix A.\hfill \QEDclosed

With the constructed surrogate functions, problem (\ref{linear_p4}) can be recast to
\begin{subequations}\label{linear_p5}
	\begin{align}
&\max_{\{\mathbf{P}_m\},\{C^c_{k,m}\}, R_\text{r}} R_\text{r} -\frac{1}{\rho}\sum^{M}_{m=1}||\mathbf{P}_m-\mathbf{FT}_m\mathbf{W}_m||^2_F,\label{ob_a5}\\
	\text{s.t.}~&\sum_{k=1}^{K}C^c_{k,m} \leq f^c_{k,m}\left(\mathbf{P}_m\right),~\forall m, \forall k,\label{ob_b5}\\
    &\sum_{m=1}^{M}\left(C^c_{k,m} +f^p_{k,m}\left(\mathbf{P}_m\right)\right)\geq R_\text{r},~\forall m, \forall k,\label{ob_c5}\\
  &\mbox{ (\ref{ob_d}), (\ref{ob_c2})}.\label{ob_d5}
	\end{align}
\end{subequations}
Given the known expansion point $\tilde{\mathbf{p}}_{k,m}$ for $\forall k$ and $\forall m$, all constraints remain convex sets. Consequently, problem (\ref{linear_p5}) is a convex optimization problem, which can be efficiently solved using standard off-the-shelf solvers. The proposed MM-based iterative algorithm is summarized in Algorithm \ref{Alg.1}.
\begin{algorithm}[t]
	\caption{MM-based iterative algorithm for solving (\ref{linear_p4})}
	\begin{algorithmic}[1]\label{Alg.1}
		\STATE Initialize $\mathbf{p}_{k,m}$ for $\forall k$ and $\forall m$. 
		\REPEAT
        \STATE Update $\tilde{\mathbf{p}}_{k,m}=\mathbf{p}_{k,m}$.
        \STATE Update surrogate $f^\tau_{k,m}\left(\mathbf{P}_m\right)$ based on equation (\ref{Overall_SCA}).
		\STATE  Solving problem (\ref{linear_p5}) to obtain optimal $\mathbf{p}_{k,m}$.
		\UNTIL{the increment of the objective value of problem  (\ref{linear_p4}) falls below a threshold.}	
		\STATE Output the optimized $\mathbf{P}_m$ and $C^c_{k,m}$ for $\forall k$ and $\forall m$.
	\end{algorithmic}
\end{algorithm}

\subsection{Subproblem w.r.t. $\mathbf{F}$ and $\{\mathbf{T}_m\}$}
The variables $\mathbf{F}$ and $\{\mathbf{T}_m\}$ only appear in the last term of the objective function and constraints (\ref{ob_e}) and (\ref{ob_f}), so problem (\ref{linear_p3}) is reduced to
\begin{subequations}\label{linear_s1}
	\begin{align}
&\min_{\mathbf{F},\{\mathbf{T}_m\}} \sum^{M}_{m=1}||\mathbf{P}_m-\mathbf{FT}_m\mathbf{W}_m||^2_F,\label{ob_sa1}\\
	\text{s.t.}~
  &\mbox{ (\ref{ob_e}), (\ref{ob_f})}.\label{ob_sb1}
	\end{align}
\end{subequations}
According to equations (\ref{dependent}) and (\ref{independent}), we observe that the analog beamfocusing $\mathbf{FT}_m$ is a full matrix, incurring coupling between the time delay of different TTDs. To decouple the interwoven variables, we introduce additional auxiliary variables $\mathbf{G}_m=\mathbf{FT}_m$ for $\forall m$. Similar to problem (\ref{linear_p2}), the penalty method is employed to migrate the equality constraint into the objective function, so the resultant new problem is
\begin{subequations}\label{linear_s2}
	\begin{align}
&\min_{\mathcal{V}_1} \sum^{M}_{m=1}\left(||\mathbf{P}_m-\mathbf{G}_m\mathbf{W}_m||^2_F+\frac{1}{\tilde{\rho}}||\mathbf{G}_m-\mathbf{FT}_m||^2_F\right),\label{ob_sa2}\\
	\text{s.t.}~
  &\mbox{ (\ref{ob_e}), (\ref{ob_f})}.\label{ob_sb2}
	\end{align}
\end{subequations}
where $\mathcal{V}_1=\left\{\mathbf{F},\{\mathbf{T}_m\},\{\mathbf{G}_m\}\right\}$ collects all optimization variables and $\tilde\rho$ is the penalty factor. When $\rho\to 0$, the equality constraint $\mathbf{G}_m=\mathbf{FT}_m$ for $\forall m$ can be protected. Currently, the introduced auxiliary matrices seem to make the problem more complex, but we will later show that they facilitate the closed-form update of analog beamfocusing. To obtain the closed-form solution, we divide optimization variables into three blocks and then utilize BCD to alternately update each block, summarized as follows.

\subsubsection{Optimization of $\{\mathbf{G}_m\}$} With fixed $\mathbf{F}$ and $\{\mathbf{T}_m\}$, problem (\ref{linear_s2}) can be decomposed into $M$ independent subproblem without any constraints, the $m$-th subproblem is given by
\begin{align}\label{linear_s3}
\min_{\mathbf{G}_m}||\mathbf{P}_m-\mathbf{G}_m\mathbf{W}_m||^2_F+\frac{1}{\tilde{\rho}}||\mathbf{G}_m-\mathbf{FT}_m||^2_F.
\end{align}
By setting the first-order derivative to zero, we get the optimal $\mathbf{G}^*_m$ with closed-form expressions, as follows
\begin{equation}
\mathbf{G}^*_m=\left(\mathbf{P}_m\mathbf{W}^H_m+\frac{1}{\tilde{\rho}}\mathbf{F}\mathbf{T}_m\right)\left(\mathbf{W}_m\mathbf{W}^H_m+\frac{1}{\tilde{\rho}}\mathbf{I}_A\right)^{-1},~\forall m.
\label{Auxiliary}
\end{equation} 

\subsubsection{Optimization of $\mathbf{F}$} Given the known $\{\mathbf{T}_m\}$ and $\{\mathbf{G}_m\}$, problem (\ref{linear_s2}) reduces to
\begin{subequations}\label{linear_s4}
	\begin{align}
&\min_{\mathbf{F}} \sum^{M}_{m=1}||\mathbf{G}_m-\mathbf{FT}_m||^2_F,\label{ob_sa4}\\
	\text{s.t.}~
  &|\mathbf{f}_{a,q,i}|=1,~\forall a, \forall q, \forall i.\label{ob_sb4}
	\end{align}
\end{subequations}
Based on equations (\ref{dependent}) and (\ref{independent}), the objective function can be recast to 
\begin{align}\label{New_obj}
&\sum^{M}_{m=1}||\mathbf{G}_m-\mathbf{FT}_m||^2_F\notag\\
=&\sum^{A}_{a=1}\sum^{Q}_{q=1}\sum^{M}_{m=1}||\mathbf{g}_{m,a,q}-\mathbf{f}_{a,q}e^{-j2\pi f_mt_{a,q}}||^2\notag\\
=&\sum^{A}_{a=1}\sum^{Q}_{q=1}\sum^{M}_{m=1}\Big(\eta-2\text{Re}\left(\mathbf{g}^H_{m,a,q}\mathbf{f}_{a,q}e^{-j2\pi f_mt_{a,q}}\right)\Big),
\end{align}
where $\mathbf{g}_{m,a,q}=\mathbf{G}_m\left(\left(q-1\right)\frac{N}{Q}+1:q\frac{N}{Q},a\right)$ and $\eta=\mathbf{g}^H_{m,a,q}\mathbf{g}_{m,a,q}+\mathbf{f}^H_{a,q}\mathbf{f}_{a,q}$. Two key observations emerge regarding the new objective function. First, due to the unit-modulus constraint on $\mathbf{f}_{a,q,i}$, one has $\mathbf{f}^H_{a,q}\mathbf{f}_{a,q}=N/Q$, so $\eta$ remains a constant. Consequently, the optimization depends solely on the second term in (\ref{New_obj}). Second, vector $\mathbf{f}_{a,q}$ for $\forall a$ and $\forall q$ has been completely decoupled and can be optimized independently. Leveraging these insights, the new objective function can be decomposed into $AQ$ independent subproblems. For $\mathbf{f}_{a,q}$, its optimization problem can be given by
\begin{subequations}\label{linear_s5}
	\begin{align}
&\max_{\mathbf{f}_{a,q}}\sum^{M}_{m=1} \text{Re}\left(\mathbf{g}^H_{m,a,q}\mathbf{f}_{a,q}e^{-j2\pi f_mt_{a,q}}\right),\label{ob_sa5}\\
	\text{s.t.}~
  &|\mathbf{f}_{a,q,i}|=1,~\forall i.\label{ob_sb5}
	\end{align}
\end{subequations}
Its optimal solution is
\begin{equation}
\mathbf{f}^*_{a,q}=\left( e^{-j\angle\sum^{M}_{m=1} \left(\mathbf{g}^H_{m,a,q}e^{-j2\pi f_mt_{a,q}}\right)}\right)^T.
\label{PS_beamfocusing}
\end{equation}

\subsubsection{Optimization of $\{\mathbf{T}_m\}$} When fixing  $\{\mathbf{F}\}$ and $\{\mathbf{G}_m\}$, problem (\ref{linear_s2}) reduces to
\begin{subequations}\label{linear_s6}
	\begin{align}
&\min_{\{\mathbf{T}_m\}} \sum^{M}_{m=1}||\mathbf{G}_m-\mathbf{FT}_m||^2_F,\label{ob_sa6}\\
	\text{s.t.}~
  &t_{a,q}\in\left[0,t_{\text{max}}\right],~\forall a,\forall q.\label{ob_sb6}
	\end{align}
\end{subequations}
Based on equation (\ref{New_obj}), variables $t_{a,q}$  can be decoupled and optimized independently. The optimization problem with respect to $t_{a,q}$ can be reformulated as
\begin{subequations}\label{linear_s7}
	\begin{align}
&\max_{t_{a,q}}\sum^{M}_{m=1} \text{Re}\left(\mathbf{g}^H_{m,a,q}\mathbf{f}_{a,q}e^{-j2\pi f_mt_{a,q}}\right),\label{ob_sa5}\\
	\text{s.t.}~
  &t_{a,q}\in\left[0,t_{\text{max}}\right].\label{ob_sb7}
	\end{align}
\end{subequations}
This constitutes a classical single-variable optimization problem over a fixed interval, which can be efficiently solved through one-dimensional search. Specifically, by defining a searching set $\mathcal{S}=\left\{0,t_{\text{max}}/(S-1),2t_{\text{max}}/(S-1),\dots,t_{\text{max}}\right\}$ with $S$ sample points, a nearly-optimal $t_{a,q}$ can be obtained as 
\begin{equation}
t^*_{a,q} = \arg\max_{t_{a,q}\in\mathcal{S}}\sum^{M}_{m=1} \text{Re}\left(\mathbf{g}^H_{m,a,q}\mathbf{f}_{a,q}e^{-j2\pi f_mt_{a,q}}\right)
\label{one-dimension}
\end{equation} 
Building upon the block-wise solutions derived above, we outline the proposed alternating optimization algorithm in Algorithm \ref{Alg.2}.
\begin{algorithm}[t]
	\caption{Alternating optimization for solving (\ref{linear_s2})}
	\begin{algorithmic}[1]\label{Alg.2}
        \STATE Initialize $\mathbf{F}$ and $\mathbf{T}_m$ for $\forall m$. Initialize $0<\alpha<1$ and $\tilde \rho$. Set the maximum tolerance $\xi_2$.
        \REPEAT
		\REPEAT
		\STATE  Update $\mathbf{G}_m$ based on equation (\ref{Auxiliary}) for $\forall m$.
        \STATE Update $\mathbf{F}$ based on equation (\ref{PS_beamfocusing}).
        \STATE Update $\mathbf{T}_m$ based on equation (\ref{one-dimension}) for $\forall m$.
		\UNTIL{ the decrement of the objective value of problem (\ref{linear_s2}) falls below a threshold. }
        \STATE Update penalty factor $\tilde \rho=\alpha \tilde \rho$.
        \UNTIL{the penalty value falls below a threshold.}
		\STATE Output the optimized $\mathbf{F}$ and $\mathbf{T}_m$ for $\forall m$.
	\end{algorithmic}
\end{algorithm}

\subsection{Subproblem w.r.t. $\{\mathbf{W}_m\}$ }
Since the digital beamfocusing $\{\mathbf{W}_m\}$ appears only in the second term of the objective function, problem (\ref{linear_p3}) simplifies to
\begin{align}\label{linear_p6}
&\min_{\{\mathbf{W}_m\} } \sum^{M}_{m=1}||\mathbf{P}_m-\mathbf{FT}_m\mathbf{W}_m||^2_F,
\end{align} 
which is a quadratic and separable problem across subcarriers over $\mathbf{W}_m$. By solving $\frac{\partial ||\mathbf{P}_m-\mathbf{FT}_m\mathbf{W}_m||^2_F}{\partial \mathbf{W}_m}=0$, the optimal  $\mathbf{W}^*_m$ is found as  
\begin{equation}
\mathbf{W}^*_m=\left(\mathbf{T}^H_m\mathbf{F}^H\mathbf{F}\mathbf{T}_m\right)^{-1}\mathbf{T}^H_m\mathbf{F}^H\mathbf{P}_m,~\forall m.
\label{Optimal_digital}
\end{equation} 

\subsection{Overall algorithm and its properties}
Based on the solutions derived for each block, we now present the complete penalty-based iterative algorithm in Algorithm~\ref{Alg.3}. Here are its crucial properties, including convergence and complexity.

\begin{itemize}
\item  \emph{Convergence}: Starting from an arbitrary feasible point, Algorithm~\ref{Alg.1} maintains at least the feasibility and performance of the previous solution at each iteration, thereby generating a monotonically non-decreasing sequence of objective values. Since the transmit rate is lower-bounded by a finite value, this monotonicity property guarantees the convergence of Algorithm~\ref{Alg.1}. As for Algorithm~\ref{Alg.2} and Algorithm~\ref{Alg.3}, the convergence is ensured by three key properties: 1) each block variable is optimally updated in every inner-loop iteration, 2) the objective function is bounded, and 3) the penalty-based approach in the outer-loop iteration has been shown to converge to a stationary point\cite{9120361}. These conditions collectively guarantee that both algorithms converge within a finite number of iterations.

\item \emph{Complexity}: The main load in Algorithm~\ref{Alg.1} comes from solving the convex problem (\ref{linear_p5}). Using a conventional interior point method, its complexity is $\mathcal O\left(N_v^{3.5}\right)$, where $N_v=MN(K+1)+MK$ is the number of variables. In contrast, Algorithm~\ref{Alg.2} is computationally efficient, as it updates variables either via closed-form solutions or one-dimensional searches. Specifically, the update procedures for $\mathbf{G}_m$, $\mathbf{F}$, and $\mathbf{T}_m$ exhibit computational complexities of $\mathcal O\left(MAN(K+1)+MA^2(NQ+N+K+1)+MA^3\right)$, $\mathcal O\left(MNA\right)$, and $\mathcal O\left(M^2NSA\right)$, respectively. Furthermore, the complexity of updating $\mathbf{W}_m$ in Algorithm~\ref{Alg.3} is in order of $\mathcal O\left(2MA^2N +MA^3+MAN(K+1)\right)$.
\end{itemize}

\begin{algorithm}[t]
	\caption{Penalty-based iterative algorithm for solving (\ref{linear_p})}
	\begin{algorithmic}[1]\label{Alg.3}
		\STATE Initialize $\mathbf{F}$, $\mathbf{W}_m$, and $\mathbf{T}_m$ for $\forall m$. Initialize $0<\alpha<1$ and $\rho$. Set the maximum tolerance $\xi_3$.
        \REPEAT
		\REPEAT
		\STATE  Update $\mathbf{P}_m$  for $\forall m$ via Algorithm~\ref{Alg.1}.
        \STATE Update $\mathbf{F}$ and $\mathbf{T}_m$  for $\forall m$ via Algorithm~\ref{Alg.2}.
        \STATE Update $\mathbf{W}_m$ based on equation (\ref{linear_p}) for $\forall m$.
		\UNTIL { the increment of the objective value of problem (\ref{linear_s2}) falls below a threshold. }
        \STATE Update penalty factor $\rho=\alpha  \rho$.
        \UNTIL{the penalty value falls below a threshold. }
		\STATE Output the optimized max-min transmit rate.
	\end{algorithmic}
\end{algorithm}

\section{Extension to sub-connected TTD-based architecture}\label{Section IV}
This section introduces a sub-connected TTD-based analog beamfocusing architecture and extends our proposed algorithm to accommodate such configurations.
\begin{figure}[tbp]
\centering
\includegraphics[scale=0.5]{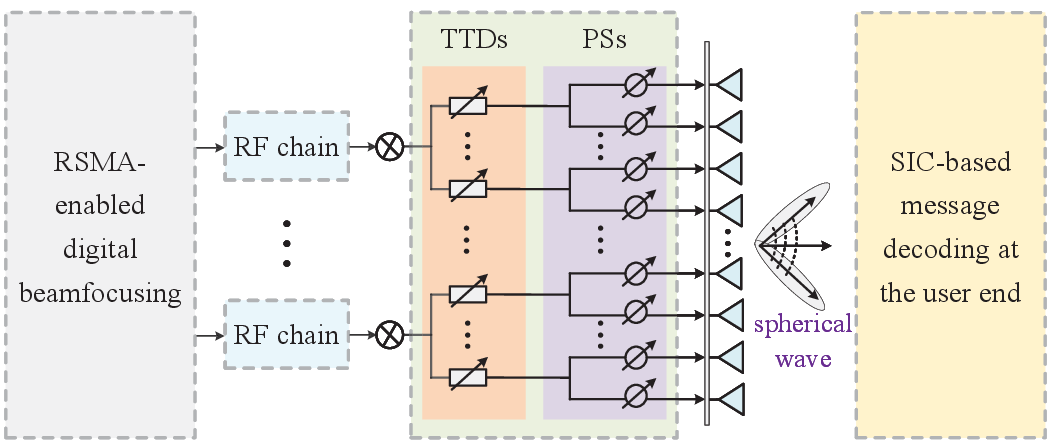}
\caption{Sub-connected TTD architecture for RSMA-enabled wideband NFC.}
\label{system_2}
\end{figure}  

In the fully-connected TTD architecture, both the number of required TTD elements and their maximum delay specifications grow linearly with the ELAA's aperture size\cite{10541333,9735144}. This scaling characteristic results in prohibitively high power consumption in TTD networks, particularly for large-aperture array scenarios. To overcome this limitation, we propose a sub-connected architecture for RSMA-enabled wideband NFC system, as shown in Fig~\ref{system_2}.  In this architecture, each RF chain is connected to a sub-array comprising $\tilde N = N/A$ antennas, while each TTD unit interfaces with a further partitioned sub-array of $\tilde N/Q$ antennas. Under this configuration, the frequency-independent analog beamfocusing matrix can be expressed as  
\begin{subequations}\label{independent_sub}
\begin{align}
& \mathbf{F} =\text{blkdiag}\left(\mathbf{F}_1,\dots,\mathbf{F}_A\right),\\
&\mathbf{F}_a= \text{blkdiag}\left(\mathbf{f}_{a,1},\dots,\mathbf{f}_{a,Q}\right),~\forall a\in\mathcal{A},
\end{align}
\end{subequations}
where $\mathbf{F}_a\in\mathbb{C}^{\tilde N\times Q}$ and $\mathbf{f}_{a,q}\in\mathbb{C}^{\tilde N/Q\times 1}$ represent the PS-based analog beamfocusing connected the $a$-th RF chain and the $q$-th TTD of this chain, respectively. 

Next, we extend our proposed penalty-based iterative algorithm to support such architectures. Importantly, the optimization of $\{\mathbf{P}_m\}$, $\{C^c_{k,m}\}$, and $\{\mathbf{W}_m\}$ maintain the same structure as in the first and third subproblems (Section \ref{Section III}), as they are decoupled of the frequency-independent analog beamfoucsing design. We therefore focus on the subprobem concerning $\mathbf{F}$ and $\{\mathbf{T}_m\}$, which can be formulated as
\begin{subequations}\label{linear_e1}
	\begin{align}
&\min_{\mathbf{F},\{\mathbf{T}_m\}} \sum^{M}_{m=1}||\mathbf{P}_m-\mathbf{FT}_m\mathbf{W}_m||^2_F,\label{ob_ea1}\\
	\text{s.t.}~
  &|\mathbf{f}_{a,q,i}|=1,~\forall a, \forall q, \forall i,\label{ob_eb1}\\  
  &t_{a,q}\in\left[0,t_{\text{max}}\right],~\forall a, \forall q.\label{ob_ec1}
	\end{align}
\end{subequations}
where $1\leq i\leq \tilde N/Q$. In contrast to the fully-connected architecture, the analog beamforming matrix $\mathbf{FT}_m$ in the sub-connected architecture exhibits a block-diagonal structure rather than being a full matrix. Leveraging this special sparse structure, we can solve problem (\ref{linear_e1}) directly without introducing any auxiliary variables. Specifically, we recast the objective function as follows:
\begin{align}\label{New_obj_2}
&\sum^{M}_{m=1}||\mathbf{P}_m-\mathbf{FT}_m\mathbf{W}_m||^2_F\notag\\
=&\sum^{A}_{a=1}\sum^{M}_{m=1}||\tilde{\mathbf{P}}_{m,a}-\mathbf{F}_{a}e^{-j2\pi f_m\mathbf{t}_{a}}\tilde{\mathbf{w}}^H_{m,a}||^2_F\notag\\
=&\tilde\eta-\sum^{A}_{a=1}\sum^{M}_{m=1}2\text{Re}\left(\tilde{\mathbf{w}}^H_{m,a}\tilde{\mathbf{P}}^H_{m,a}\mathbf{F}_{a}e^{-j2\pi f_m\mathbf{t}_{a}}\right)\notag\\
=&\tilde\eta-\sum^{A}_{a=1}\sum^{Q}_{q=1}\sum^{M}_{m=1}2\text{Re}\left(\mathbf{\psi}^H_{m,a,q}\mathbf{f}_{a,q}e^{-j2\pi f_mt_{a,q}}\right),
\end{align}
where 
\begin{align}\label{New_obj_3}
&\tilde{\mathbf{P}}_{m,a}=\mathbf{P}_m\left((a-1)\tilde N+1:a\tilde N,:\right),\notag\\
&\tilde{\mathbf{w}}_{m,a} = \mathbf{W}^H_m(:,a),\notag\\
&\tilde \eta \overset{(a)}=\sum^{A}_{a=1}\sum^{M}_{m=1}\left(\tilde N\tilde{\mathbf{w}}^H_{m,a}\tilde{\mathbf{w}}_{m,a}+\text{Tr}\left(\tilde{\mathbf{P}}_{m,a}\tilde{\mathbf{P}}^H_{m,a}\right)\right),\notag\\
&\mathbf{\Psi}_{m,a}=\tilde{\mathbf{P}}_{m,a}\tilde{\mathbf{w}}_{m,a},\notag\\
&\mathbf{\psi}_{m,a,q}=\mathbf{\Psi}_{m,a}\left((q-1)\tilde N/Q+1:q\tilde N/Q,1\right),
\end{align}
where equation (a) holds due to the unit-modulus constraint on $\mathbf{f}_{a,q,i}$.
From equation (\ref{New_obj_2}), it can be seen that the original optimization problem admits a decomposition into $AQ$ independent subproblems. In particular, the joint optimization of $\mathbf{f}_{a,q}$ and $t_{a,q}$ can be formulated as:
\begin{subequations}\label{linear_e2}
	\begin{align}
&\max_{\mathbf{f}_{a,q},t_{a,q}}\sum^{M}_{m=1}\text{Re}\left(\mathbf{\psi}^H_{m,a,q}\mathbf{f}_{a,q}e^{-j2\pi f_mt_{a,q}}\right),\label{ob_ea2}\\
	\text{s.t.}~
  &|\mathbf{f}_{a,q,i}|=1,~\forall i.\label{ob_eb2}\\
  &t_{a,q}\in\left[0,t_{\text{max}}\right].\label{ob_ec2}
	\end{align}
\end{subequations}
With fixed $t_{a,q}$, the optimal $\mathbf{f}^*_{a,q}$ is given by
\begin{equation}
\mathbf{f}^*_{a,q}=\left(e^{-j\angle\sum^{M}_{m=1} \left(\mathbf{\psi}^H_{m,a,q}e^{-j2\pi f_mt_{a,q}}\right)}\right)^T
\label{PS_beamfocusing_2}
\end{equation}
With fixed $\mathbf{f}_{a,q}$, adopting one-dimensional search, the nearly-optimal $t^*_{a,q}$ is 
\begin{equation}
t^*_{a,q} = \arg\max_{t_{a,q}\in\mathcal{S}}\sum^{M}_{m=1} \text{Re}\left(\mathbf{\psi}^H_{m,a,q}\mathbf{f}_{a,q}e^{-j2\pi f_mt_{a,q}}\right)
\label{one-dimension_2}
\end{equation} 

The proposed algorithm for the sub-connected architecture is summarized in Algorithm~\ref{Alg.4}. Its convergence analysis adheres to the same methodological framework established in Algorithm~\ref{Alg.3}. In each iteration, the computational complexity of updating $\mathbf{P}_m$ and $\mathbf{W}_m$ remains identical to that of Algorithm~\ref{Alg.1}. Meanwhile, the complexities associated with updating $\mathbf{F}$ and $\mathbf{T}_m$ are of the order of $\mathcal O\left(M\left(N^2\left(K+1\right)^2+N\right)\right)$ and $\mathcal O\left(MN^2S/AQ\right)$, respectively.

\begin{algorithm}[t]
	\caption{Penalty-based iterative algorithm for sub-connected architecture}
	\begin{algorithmic}[1]\label{Alg.4}
		\STATE Initialize $\mathbf{F}$, $\mathbf{W}_m$, and $\mathbf{T}_m$ for $\forall m$. Initialize $0<\alpha<1$ and $\rho$. Set the maximum tolerance $\xi_4$.
        \REPEAT
		\REPEAT
		\STATE  Update $\mathbf{P}_m$  for $\forall m$ via Algorithm~\ref{Alg.1}.
        \REPEAT
        \STATE Update $\mathbf{F}$ based on equation (\ref{PS_beamfocusing_2}).
        \STATE Update $\mathbf{T}_m$ based on equation (\ref{one-dimension_2}) for $\forall m$.
        \UNTIL{the decrement of the objective value of problem (\ref{linear_e1}) falls below a threshold.}
        \STATE Update $\mathbf{W}_m$ based on equation (\ref{Optimal_digital}) for $\forall m$.
		\UNTIL{the increment of the objective value of problem (\ref{linear_s2}) falls below a threshold.}
        \STATE Update penalty factor $\rho=\alpha  \rho$.
        \UNTIL{the penalty value falls below a threshold.}
		\STATE Output the optimized max-min transmit rate.
	\end{algorithmic}
\end{algorithm}

\section{Simulation results}\label{Section V}
This section evaluates our proposed transmit scheme and algorithms. Unless otherwise specified, the simulation parameters are configured as follows: The BS is equipped with $N=128$ antennas with a half-wavelength spacing. The central frequency, system bandwidth, and number of subcarriers are set to $f_c=30$~GHz, $B=10$~GHz, and $M=10$, respectively. The RF chains and scatters associated with each user are set to $A=8$ and $L_k=2$, respectively. Each RF chain is connected to $Q=16$ TTDs, each with a maximum time delay of $t_{\max}=N/(2f_c)=2.13$ nanosecond (ns)\cite{9735144}.  $K=4$ users and scatters are randomly generated within the distance from $10$~m to $20$~m. The maximum transmit power at the BS and background noise spectral density are $P_{\text{th}}=20$~dBm and $\sigma^2=-174$~dBm/Hz, respectively. The penalty factors are set to $\rho=\tilde\rho=10^2$, with a reduction factor of $\alpha=0.5$. For the one-dimensional search, we set $S=10^3$ steps. These parameter settings are primarily sourced from \cite{10579914,10587118}.

Over $100$ independent channel realizations, we test our proposed fully-connected and sub-connected TTD-based hybrid beamfocusing architectures for RSMA-enabled wideband NFC (labeled as {\bf{RSMA-FHB-near}} and {\bf{RSMA-SHB-near}}). To assess its performance comprehensively, the proposed scheme is compared against four relevant benchmarks, which are detailed as follows:
\begin{itemize}
\item {\bf{RSMA-FDB-near}}: The BS adopts a full digital beamfocusing structure, where a dedicated RF chain drives each antenna. This benchmark serves as the theoretical upper bound on the performance of our transmit scheme. 
\item {\bf{RSMA-PS-near}}: This uses analog beamfocusing for only a frequency-independent PS network. The optimization problem for this case is solved by setting $t_{a,q}=0$ for $\forall a$ and $\forall q$.
\item {\bf{SDMA-FHB-near}}: This benchmark adopts the fully-connected TTD-based hybrid beamfocusing architecture but employs the conventional SDMA technique. The message for each user is encoded into a user-specific stream, and receivers treat interference as noise. This means that this baseline disables the common stream, \emph{i.e.}, $\mathbf{w}_{0,m}=\mathbf{0}$.
\item {\bf{RSMA-FHB-far}}: This one adopts the far-field channel model, where the array response vector in equation (\ref{Channel}) is updated to
\begin{equation}
\mathbf{a}_{\text{far}}\left(\theta_k,f_m\right)= \left[e^{j\frac{2\pi f_m}{c}\frac{1-N}{2}d\sin\theta_k},\dots,e^{j\frac{2\pi f_m}{c}\frac{N-1}{2}d\sin\theta_k}\right]^T
\label{Far-Channel}
\end{equation}
All other parameters remain identical to ensure comparison fairness.
\end{itemize}

\begin{figure}[tbp]
	\centering
	\includegraphics[scale=0.55]{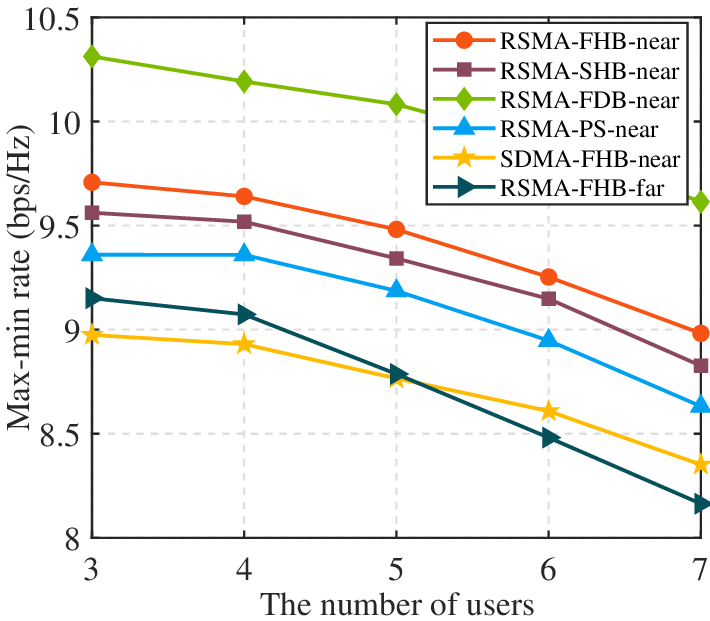}
	\caption{Max-min rate versus the number of users.}
	\label{user}
\end{figure}
Fig.~\ref{user} illustrates the max-min rate versus the number of users. The simulation result highlights four key advantages of the proposed transmit scheme over competing benchmarks. 
\begin{enumerate}
    \item \emph{Near-field beamfocusing superiority:} The proposed near-field beamfocusing surpasses far-field beamforming, delivering a gain of $0.55$~bps/Hz for $K=4$ users. Moreover, the performance gaps gradually widen as user density increases. This improvement arises from the near-field beams' precise spatial energy concentration capability, which substantially reduces inter-user interference through effective beam leakage suppression.
    \item \emph{Superior anti-interference capability:} Although all considered schemes exhibit degraded max-min rate under increasing user load, RSMA consistently maintains a $0.75$~bps/Hz rate gain over SDMA. This result demonstrates RSMA's superior interference management capabilities and the inherent limitations of near-field beamfocusing in completely suppressing multi-user interference.
    \item \emph{Spatial wideband effect compensation:}  Our TTD-based beamfocusing architectures achieve significant improvements over PS-based analog approaches, with the fully-connected and sub-connected configurations delivering $0.3$~bps/Hz and $0.17$~bps/Hz gains, respectively.
    \item \emph{Comparable to full digital beamfocusing:} Our proposed TTD-based HAD beamfocusing achieves comparable performance to a fully-digital architecture while requiring only eight RF chains, demonstrating its hardware efficiency.
\end{enumerate}

\begin{figure}[tbp]
	\centering
	\includegraphics[scale=0.55]{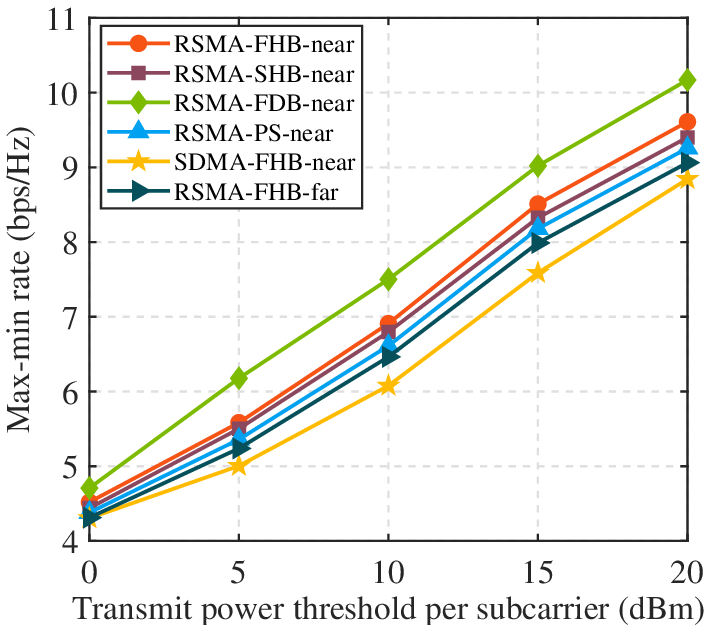}
	\caption{Max-min rate versus transmit power per subcarrier.}
	\label{power}
\end{figure}
Fig.~\ref{power} illustrates the max-min rate as a function of the transmit power threshold per subcarrier. Several key observations can be made. First, all schemes exhibit improved communication rates as the transmit power increases. Notably, RSMA achieves a $5.2$~bps/Hz gain, compared to $4.4$~bps/Hz for SDMA, as $P_{\text{th}}$ increases from $0$~dBm to $20$~dBm. This advantage stems from RSMA's enhanced interference management and the larger optimization space enabled by higher power levels. Second, the proposed transmit scheme consistently outperforms both PS-based analog beamfocusing and conventional far-field beamforming, with performance gains becoming more pronounced at higher power levels. Third, the proposed hybrid beamfocusing architectures closely approach the performance of the fully digital architecture while requiring 16 times fewer RF chains, demonstrating a favorable trade-off between complexity and performance.

\begin{figure}[tbp]
	\centering
	\includegraphics[scale=0.55]{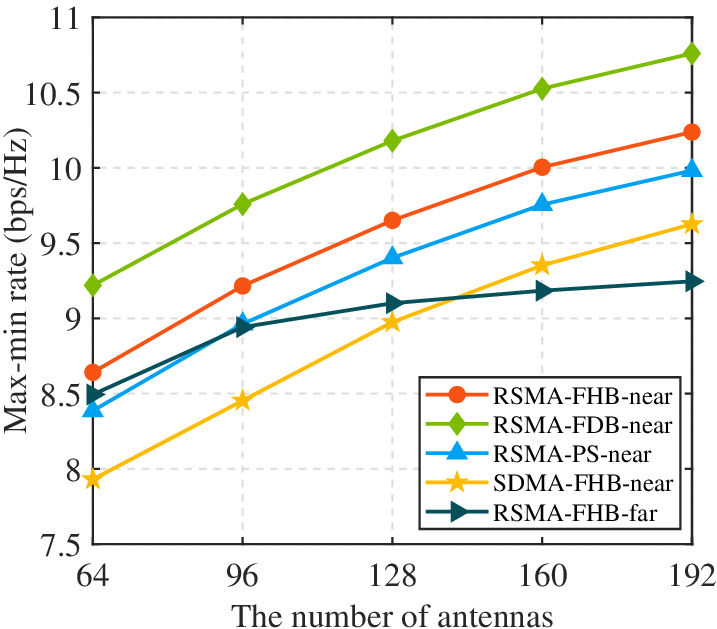}
	\caption{Max-min rate versus the number of antennas.}
	\label{antenna}
\end{figure}
Fig.~\ref{antenna} investigates the impact of the number of transmit antennas on the max-min rate. As expected, all schemes benefit from increasing antenna counts due to improved spatial multiplexing gains and greater spatial degrees of freedom. However, our proposed scheme achieves notable performance gains of $0.65$~bps/Hz and $0.2$~bps/Hz over conventional SDMA and PS-based HAD beamfocusing, respectively. Additionally, the performance gap between near-field beamfocusing and far-field beamforming widens as the antenna array grows. Most notably, the results reveal a fundamental performance crossover: far-field RSMA experiences significant degradation and is ultimately surpassed by near-field SDMA when $N \geq 128$. This highlights the superior interference suppression and signal enhancement capabilities of near-field beamfocusing, particularly in large-array regimes.

\begin{figure}[tbp]
	\centering
	\includegraphics[scale=0.55]{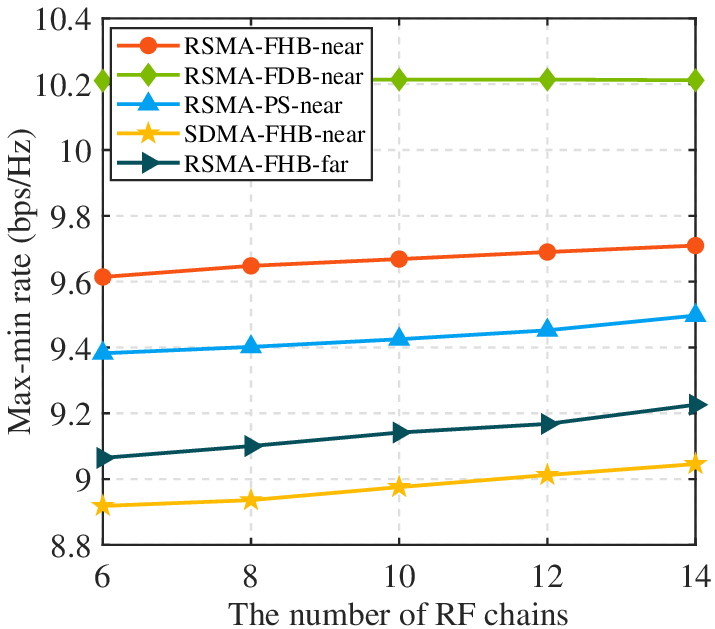}
	\caption{Max-min rate versus the number of RF chains.}
	\label{radio_frequency}
\end{figure}
Fig.~\ref{radio_frequency} presents the max-min rate versus the number of RF chains. The fully-digital beamfocusing architecture maintains a constant communication rate, establishing an upper-bound performance benchmark. The performance gap between the fully-digital and TTD-based HAD architectures exhibits a diminishing trend with increasing RF chains, narrowing to just $0.5$~bps/Hz at $N=14$. This convergence performance gap stems from the improved interference mitigation capability enabled by the expanded digital beamforming dimensions. However, with an identical number of RF chains, our proposed transmit scheme always surpasses the other three benchmarks. our proposed scheme demonstrates consistent performance superiority over all three benchmarks under identical RF chain configurations, highlighting its practical advantages in NFC applications.

\begin{figure}[H]
	\centering
	\subfigure{
		\begin{minipage}[H]{0.45\linewidth}
			\centering\includegraphics[width = 1.53in]{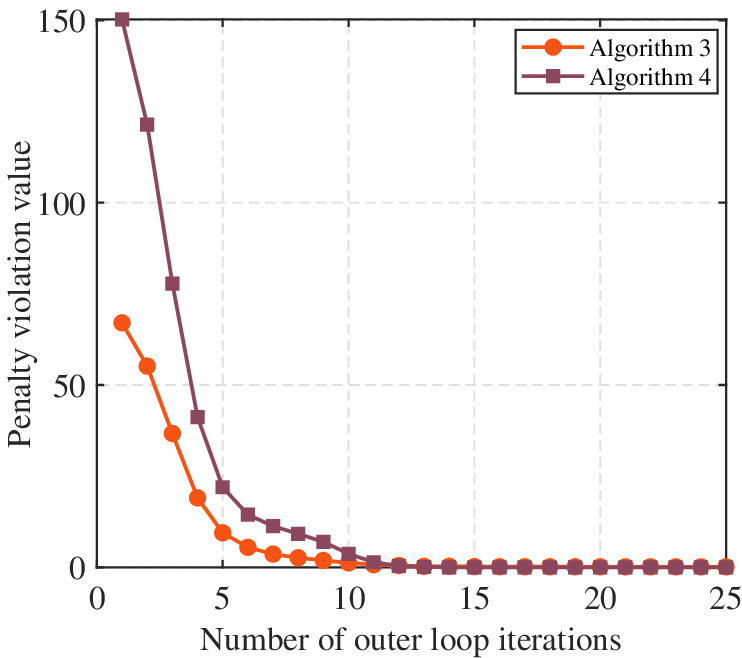}
		\end{minipage}
	}
	\subfigure{
		\begin{minipage}[H]{0.48\linewidth}
			\centering\includegraphics[width = 1.53in]{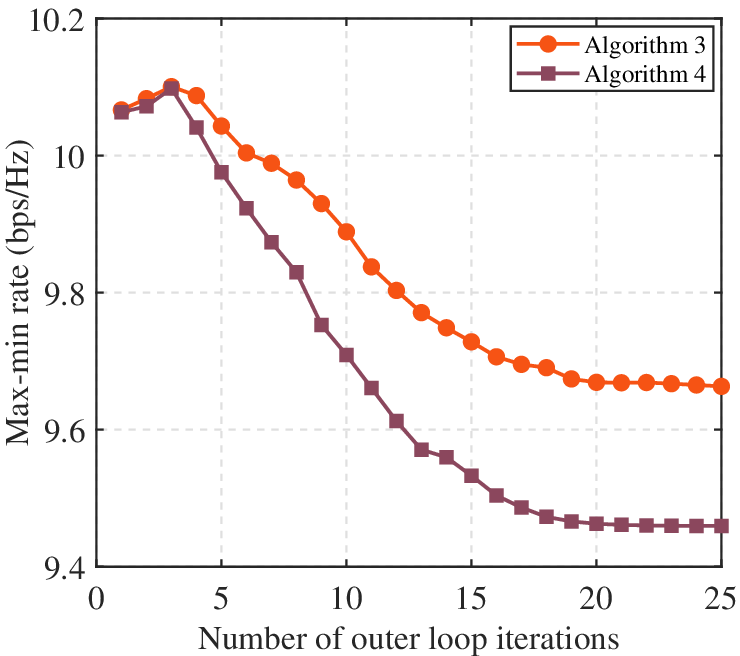}
		\end{minipage}
	}
	\centering
	\caption{Convergence behavior of the proposed algorithms.}
	\label{convergence}
\end{figure}
Fig.~\ref{convergence} presents the convergence behaviors, evaluating the penalty violation value (left) and the max-min rate (right) as functions of the outer loop iteration index. As can be seen, the equality constraint violation value $\sum^{M}_{m=1}||\mathbf{P}_m-\mathbf{FT}_m\mathbf{W}_m||^2_F$ decreases rapidly with increasing outer-loop iterations, ensuring exact satisfaction of  $\mathbf{P}_m=\mathbf{FT}_m\mathbf{W}_m$ for $\forall m$. The max-min rate exhibits a monotonic decreasing convergence behavior, rather than increasing monotonically. This counterintuitive phenomenon arises from the initial relaxation of equality constraints, which temporarily enlarges the feasible solution space and enables higher achievable rates during early iterations. As the optimization progresses, the constraints gradually tighten to converge to the globally optimal feasible solution.

\section{Conclusion}\label{Section VI}
Wideband NFC systems face significant challenges of hardware complexity and spatial wideband effects. To address these issues, this paper proposes an RSMA-enabled transmission scheme for wideband NFC, incorporating TTD-based hybrid beamfocusing architectures. We jointly optimize frequency-dependent and frequency-independent analog beamfocusing, digital beamfocusing, and common rate allocation to maximize the minimum user rate. To tackle the resulting non-convex optimization problem, we develop a penalty-based iterative algorithm that partitions the variables into three blocks and applies BCD for alternating optimization. The algorithm is further extended to accommodate sub-connected TTD-based analog architectures.
Extensive simulations demonstrate that the proposed scheme effectively mitigates spatial wideband effects and achieves performance close to that of fully digital beamfocusing. Moreover, it significantly outperforms traditional SDMA and far-field wideband communication strategies.

This work also opens several promising research directions. First, RSMA’s inherent robustness makes it a strong candidate for handling imperfect CSI, which is notably harder to acquire in near-field scenarios. Second, the dual-dimensional spatial resolution enabled by spherical wavefronts offers potential for simultaneous angle and distance estimation. This motivates further exploration of RSMA-enabled near-field wideband ISAC systems.

\appendices
\section{Proof of Claim 1}
This appendix demonstrates that the constructed surrogate for $R^c_{k,m}$ satisfies both properties outlined in Claim 1. The same analytical framework can be directly extended to $R^p_{k,m}$. To provide rigorous proof, we first rewrite  $\log\left(1+\gamma^c_{k,m}\right)$ as
\begin{align}
\log\left(1+\frac{S^c_{k,m}}{I^c_{k,m}}\right)=-\log\left(1-\left(\mathbf{u}^c_{k,m}\right)^H\mathbf{h}^H_{k,m}\mathbf{p}_{0,m}\right),
\label{eq:S/I}
\end{align}
where $\mathbf{u}^c_{k,m}=\left(T^{c}_{k,m}\right)^{-1}\mathbf{h}^H_{k,m}\mathbf{p}_{0,m}$.
We observe that the right-hand-side of (\ref{eq:S/I}) is a convex function over $v^c_{k,m}=1-\left(\mathbf{u}^c_{k,m}\right)^H\mathbf{h}^H_{k,m}\mathbf{p}_{0,m}$. Thus, by employing first-order Taylor expansion at ${v}^c_{k,m}=\tilde{v}^c_{k,m}=1-\left(\tilde{\mathbf{u}}^c_{k,m}\right)^H\mathbf{h}^H_{k,m}\tilde{\mathbf{p}}_{0,m}$, we can obtain its lower-bound, which is given by
\begin{align}
\log\left(1+\frac{S^c_{k,m}}{I^c_{k,m}}\right)\geq-\log\left(\tilde{v}^c_{k,m}\right)-\frac{v^c_{k,m}}{\tilde{v}^c_{k,m}\ln 2}+\frac{1}{\ln 2}.
\label{eq:A2}
\end{align}
We then define an auxiliary parameter $w^c_{k,m}$, where
\begin{align}
w^c_{k,m}\overset{(a)}=&v^c_{k,m}+{\left(\tilde{\mathbf{u}}^c_{k,m}-\mathbf{u}^c_{k,m}\right)}^H T^{c}_{k,m}\left(\tilde{\mathbf{u}}^c_{k,m}-\mathbf{u}^c_{k,m}\right)\\ 
\overset{(b)}=&\left(\tilde{\mathbf{u}}^c_{k,m}\right)^HT^{c}_{k,m}\tilde{\mathbf{u}}^c_{k,m}- 2\text{Re}\left(\tilde{\mathbf{u}}^c_{k,m}\mathbf{h}^H_{k,m}\mathbf{p}_{0,m}\right)+1.\notag  
\end{align}
where equation $(b)$ holds since $T^{c}_{k,m}\in\mathbb{C}^{1\times 1}$ and $\mathbf{u}^c_{k,m}\in\mathbb{C}^{1\times 1}$ are  not multi-dimensional vectors or matrices. The equation $(a)$ indicate $w^c_{k,m}\geq v^c_{k,m}$, yielding
\begin{align}
\log\left(1+\frac{S^c_{k,m}}{I^c_{k,m}}\right)\geq&-\log\left(\tilde{v}^c_{k,m}\right)-\frac{w^c_{k,m}}{\tilde{v}^c_{k,m}\ln 2}+\frac{1}{\ln 2}\notag\\
=&f^c_{k,m}\left(\mathbf{P}_m\right).
\label{eq:A3}
\end{align}

Now, we prove the equation in (\ref{eq:A3}) holds at $\mathbf{p}_{k,m}=\tilde{\mathbf{p}}_{k,m}$ for $\forall k$. When $\mathbf{p}_{k,m}=\tilde{\mathbf{p}}_{k,m}$ for $\forall k$, we have $w^c_{k,m}=v^c_{k,m}$ and $\tilde v^c_{k,m}=v^c_{k,m}$. Based on this insight, we can derive
\begin{align}
f^c_{k,m}\left(\tilde{\mathbf{P}}_m\right)=&-\log\left(\tilde{v}^c_{k,m}\right)-\frac{w^c_{k,m}}{\tilde{v}^c_{k,m}\ln 2}+\frac{1}{\ln 2}\notag\\
=&-\log\left(\tilde{v}^c_{k,m}\right)-\frac{v^c_{k,m}}{\tilde{v}^c_{k,m}\ln 2}+\frac{1}{\ln 2}\notag\\
=& \log\left(1+\frac{S^c_{k,m}}{I^c_{k,m}}\Bigg|_ {\mathbf{p}_{k,m}=\tilde{\mathbf{p}}_{k,m}}\right).
\end{align}
We thus prove the minorization property in Claim 1.

From  the derivative rule, we can derive  
\begin{align}\label{par}
\frac{\partial{w^c_{k,m}}}{\partial{\mathbf{p}_{k,m}}} =&\frac{\partial{v^c_{k,m}}}{\partial{\mathbf{p}_{k,m}}} +\frac{\partial{{\left(\tilde{\mathbf{u}}^c_{k,m}-\mathbf{u}^c_{k,m}\right)}^H T^{c}_{k,m}} }{\partial{\mathbf{p}_{k,m}}} \left(\tilde{\mathbf{u}}^c_{k,m}-\mathbf{u}^c_{k,m}\right)\notag\\+&{\left(\tilde{\mathbf{u}}^c_{k,m}-\mathbf{u}^c_{k,m}\right)}^H T^{c}_{k,m}\frac{\partial{\left(\tilde{\mathbf{u}}^c_{k,m}-\mathbf{u}^c_{k,m}\right)} } {\partial{\mathbf{p}_{k,m}}}
\end{align}
and 
\begin{align}\label{par_2}
\frac{\partial{f^c_{k,m}\left(\mathbf{P}_m\right)}}{\partial{\mathbf{p}_{k,m}}} = -\frac{1}{\tilde{v}^c_{k,m}\ln 2}\frac{\partial{w^c_{k,m}}}{\partial{\mathbf{p}_{k,m}}}
\end{align}
Then, based on equation (\ref{eq:S/I}), the partial derivative of the original function can be given by
\begin{equation}
\frac{\partial{\log\left(1+\frac{S^c_{k,m}}{I^{c}_{k,m}}\right)}}{\partial{\mathbf{p}_{k,m}}}=-\frac{\partial{\log v^c_{k,m}}}{\partial{\mathbf{p}_{k,m}}}=-\frac{1}{v^c_{k,m}\ln 2}\frac{\partial{v^c_{k,m}}}{\partial{\mathbf{p}_{k,m}}}.
\label{partial_2}
\end{equation}

Combining (\ref{par}), (\ref{par_2}), and (\ref{partial_2}) yields
\begin{align}
&\frac{\partial{f^c_{k,m}\left(\mathbf{P}_m\right)}}{\partial{\mathbf{p}_{k,m}}}\Big|_{\mathbf{p}_{k,m} = \tilde{\mathbf{p}}_{k,m}}=-\frac{1}{\tilde{v}^c_{k,m}\ln 2}\frac{\partial{v^c_{k,m}}}{\partial{\mathbf{p}_{k,m}}}\Big|_{\mathbf{p}_{k,m} = \tilde{\mathbf{p}}_{k,m}}\notag\\&=\frac{\partial{\log\left(1+\frac{S^c_{k,m}}{I^{c}_{k,m}}\right)}}{\partial{\mathbf{p}_{k,m}}}\Big|_{\mathbf{p}_{k,m} = \tilde{\mathbf{p}}_{k,m}}.
\end{align}
which proves the gradient consistency of Claim 1.

	\ifCLASSOPTIONcaptionsoff
	\newpage
	\fi
	
	\bibliographystyle{IEEEtran}
	\bibliography{references}
	
\end{document}